\documentclass[11pt]{article}
\usepackage[margin=1in]{geometry}
\usepackage{amsmath, amssymb, amsthm}
\usepackage{booktabs}
\usepackage{graphicx}
\usepackage{hyperref}

\newtheorem{theorem}{Theorem}
\newtheorem{assumption}{Assumption}
\newtheorem{remark}{Remark}

\title{Funding-Aware Optimal Market Making for Perpetual DEXs}
\author{Nam Anh Le\\
National Economics University, Vietnam\\
\texttt{me@namanhle.com}}
\date{}

\begin{document}
\maketitle

\begin{abstract}
This paper studies optimal liquidity provision for perpetual contracts
when the funding rate is a stochastic state variable. The core extension to
classical market making is the coupling between inventory and funding payments:
inventory creates both mark-to-market exposure and a state-dependent funding
cash flow. A reduced inventory--funding control problem is formulated, solved
with a monotone finite-difference Hamilton-Jacobi-Bellman scheme, and bid and
ask quote offsets are recovered from discrete inventory value differences.
Funding is calibrated on Hyperliquid ETH, BTC, and SOL perpetual data. Gaussian OU
funding is retained as a tractable diffusion baseline, while OU-plus-jump
diagnostics document the heavy-tailed funding innovations that should enter a
future extension. In 100-seed holdout simulations under two official-fill proxy
calibrations, the funding-aware HJB improves mean ETH/BTC performance while
lowering inventory RMS relative to classical Avellaneda--Stoikov. SOL gains are
positive versus unscaled AS but are not a Pareto improvement once a risk-scaled
AS diagnostic is included.
\end{abstract}

\section{Introduction}

Perpetual futures are among the most actively traded crypto derivatives, and
their funding mechanism creates a source of risk that is absent from classical
spot-market market making. A liquidity provider who accumulates a long
perpetual inventory is not only exposed to movements in the mark price; the
position also receives or pays periodic funding as the market moves between
long- and short-pressure regimes. Funding is persistent enough to be a useful
state variable, but it also exhibits abrupt jumps and heavy-tailed innovations.
This makes it natural to ask how a market maker should quote when inventory is
jointly a price-risk exposure and a funding-cash-flow exposure.

Classical optimal market-making models, beginning with the
Avellaneda--Stoikov framework and the stochastic-control treatment in
Cartea, Jaimungal, and Penalva, focus on the trade-off between spread capture
and inventory risk \cite{avellaneda2008high,cartea2015algorithmic}. In those
models, inventory is costly because the terminal mark-to-market value is risky
and because an inventory penalty discourages large positions. Funding payments
change the sign and magnitude of the inventory incentive. A long position can
be desirable when expected funding is negative and costly when expected funding
is positive; the analogous statement holds with signs reversed for short
inventory. The market maker therefore needs a quote rule that reacts to the
funding state rather than a rule that only widens quotes as inventory grows.

The motivation is related to two strands of recent DeFi research. First,
automated-market-maker work on predictable loss, execution, and LVR studies how
liquidity providers lose value to informed or arbitrage flow
\cite{milionis2022lvr,cartea2023predictable,
cartea2025execution}. That literature is primarily about constant-function or
concentrated-liquidity mechanisms. The present paper instead studies an
order-book-style quote policy for perpetual contracts. Second, recent
perpetual-contract control work studies optimal liquidation when funding enters
wealth dynamics \cite{donnelly2026liquidation}. The present decision problem is
different: the market maker does not start with a fixed position to unwind, but
chooses bid and ask offsets that endogenously create and remove inventory
through point-process fills.

This paper makes three contributions. First, it formulates a funding-aware
market-making problem in which cash dynamics contain a state-dependent funding
term. The key object is the cash funding amount per unit inventory, denoted
$f_t$. In the data, Hyperliquid reports a fractional funding rate $F_t$; the
control state uses the cash-scaled quantity $f_t\approx S_tF_t$ so that the
funding term has the same cash units as spread capture. Second, the paper
solves the reduced inventory--funding control problem using a finite-difference
Hamilton-Jacobi-Bellman (HJB) scheme. The numerical scheme uses an upwind
birth--death approximation of the funding diffusion and enforces a CFL
condition. Quote offsets are recovered from neighboring inventory values in the
HJB table. Third, the paper calibrates the funding and fill components on
Hyperliquid ETH, BTC, and SOL data and evaluates the resulting policy against
classical funding-unaware Avellaneda--Stoikov (AS), a risk-scaled AS
diagnostic, and a practical risk-calibrated benchmark.

The empirical claim is intentionally narrow. The diffusion HJB improves mean
ETH/BTC holdout performance while lowering inventory RMS in the final 100-seed
experiments. This result survives a second official-fill calibration in which
the fill model is changed from a quote-size-aware minute-volume hit model to a
minute-hit model. The SOL result is weaker from a risk-adjusted perspective:
the HJB earns more than unscaled AS, but a risk-scaled AS diagnostic earns much
more, so SOL should be interpreted as a leverage or risk-appetite effect rather
than a Pareto improvement. The backtest is also not a queue-position simulator.
It is a controlled comparison under official-fill proxy calibrations.

The paper is conservative about the funding model. A Gaussian OU process gives
a tractable mean-reverting state for the HJB, and its estimated half-lives are
economically short enough to matter over the control horizon. However,
standardized residuals are heavy-tailed and an OU-plus-jump transition model
substantially improves likelihood for every asset. The resulting interpretation
is therefore baseline-versus-extension: this paper studies a diffusion HJB
baseline and treats jump funding as the most important theoretical extension,
not as a solved component of the current control equation.

The remainder of the paper is organized as follows. Section~\ref{sec:related}
reviews the related literature and novelty boundary.
Section~\ref{sec:prelim} sets up the stochastic-control notation and the
verification/convergence route for the finite-difference scheme.
Section~\ref{sec:model} defines the market-making model and the funding-aware
objective. Section~\ref{sec:hjb} derives the reduced HJB, quote recovery
formulas, and monotone discretization. Section~\ref{sec:calibration}
calibrates funding and fill parameters. Section~\ref{sec:backtest} reports the
cross-asset backtest and robustness checks. The final section discusses
limitations and extensions.

\section{Related Literature}
\label{sec:related}

The paper sits at the intersection of four literatures: classical market
microstructure, stochastic-control market making, automated market makers, and
perpetual-contract funding. This section is deliberately explicit about the
boundary between those literatures because the contribution is not a new
general theory of liquidity provision. It is a narrower control problem: an
order-book-style market maker in a perpetual contract, with funding as an
observed stochastic state and with quotes chosen through an inventory--funding
HJB.

\subsection{Market microstructure and dealer inventory}

The economic mechanism begins with the classical view that bid--ask spreads
compensate liquidity suppliers for transaction costs, adverse selection, order
processing, and inventory risk. Demsetz \cite{demsetz1968cost} frames the
spread as the cost of immediacy. Ho and Stoll \cite{ho1981optimal} make dealer
inventory central: a dealer who absorbs order flow changes quotes to manage
the risk of holding inventory. Glosten and Milgrom \cite{glosten1985bid}
emphasize adverse selection when some traders are informed, while Kyle
\cite{kyle1985continuous} gives a continuous-auction benchmark for strategic
informed trading and price impact. The textbook treatments of O'Hara
\cite{ohara1995market}, Harris \cite{harris2003trading}, and Hasbrouck
\cite{hasbrouck2007empirical} organize these mechanisms into the broader
market-microstructure language used by empirical trading papers.

This paper uses the inventory-risk part of that tradition but changes the
source of inventory value. In a spot instrument, inventory is costly mainly
because the terminal mark-to-market value is risky and because adverse
selection can make filled quotes stale. In a perpetual contract, inventory also
creates a funding cash flow. A long inventory can either pay or receive funding
depending on the sign of the funding rate; a short inventory has the opposite
exposure. Thus the inventory variable has two economic channels: mark-to-market
risk and funding carry. The model below isolates the second channel by replaying
historical prices rather than introducing a predictive price drift.

The optimal-execution literature is relevant because it provides the standard
continuous-time control language for trading under inventory and price impact.
Almgren and Chriss \cite{almgren2001optimal} and Bertsimas and Lo
\cite{bertsimas1998optimal} study liquidation/execution schedules rather than
two-sided quoting, but their separation of trading reward, risk penalty, and
dynamic control is conceptually close to the objective used here. Their agent
begins with an exogenous position to execute. The agent in this paper creates
and unwinds inventory endogenously through bid and ask fills.

\subsection{Optimal market making}

The direct ancestor of the model is Avellaneda and Stoikov
\cite{avellaneda2008high}. In that framework, a market maker chooses bid and
ask distances, order arrivals follow quote-dependent Poisson intensities, and
inventory aversion skews quotes away from further accumulation. Gueant,
Lehalle, and Fernandez-Tapia \cite{gueant2013dealing} sharpen this line by
deriving tractable reduced equations and asymptotic quote formulas. Cartea,
Jaimungal, and Penalva \cite{cartea2015algorithmic} provide the broader
stochastic-control development and connect market making, execution, and
high-frequency trading objectives. Guilbaud and Pham
\cite{guilbaud2013optimal} study optimal high-frequency trading with both
limit and market orders, giving a useful benchmark for models in which
inventory is discrete and controls act through order-arrival intensities.
Fodra and Labadie \cite{fodra2013high} introduce directional bets into a
market-making setting, which is conceptually related to allowing the quote
policy to react to a predictable state variable.

Limit-order-book modeling is another input. Cont, Stoikov, and Talreja
\cite{cont2010stochastic} model order-book queues as stochastic processes, and
the survey by Gould et al. \cite{gould2013limit} documents the empirical
structure of order books, including order placement, cancellation, and queue
dynamics. The present backtest does not claim queue-position realism. It uses
official Hyperliquid crossed fills and L2 snapshots to calibrate an effective
quote-fill proxy. That choice keeps the policy comparison feasible, but it
also limits the empirical interpretation. A queue-aware simulator would be the
right next layer before making a trading-production claim.

Relative to this market-making literature, the new state variable is funding.
Classical AS-style models can be recovered by setting funding to zero and
collapsing the funding grid. With funding active, the value difference between
neighboring inventory states depends on both inventory and funding. This is why
the numerical quote recovery in Section~\ref{sec:hjb} naturally skews bids and
asks as the funding state changes.

\subsection{Stochastic control and numerical HJB methods}

The theoretical backbone is the dynamic-programming approach to stochastic
control. Fleming and Soner \cite{fleming2006controlled}, Pham
\cite{pham2009continuous}, and Touzi \cite{touzi2013optimal} give the standard
references for controlled diffusions, verification, and HJB equations in
financial applications. The present paper does not prove a new stochastic-
control theorem. Instead, it states the regularity and admissibility conditions
under which the usual verification argument applies, and it uses those
conditions to motivate a finite computational domain with bounded controls.

Because the implemented policy is numerical, viscosity convergence matters more
than a smooth closed-form derivation. Crandall and Lions
\cite{crandall1983viscosity} introduced the viscosity-solution framework for
Hamilton-Jacobi equations, and Crandall, Ishii, and Lions
\cite{crandall1992users} provide the standard user's guide. Kushner and Dupuis
\cite{kushner2001numerical} develop Markov-chain approximation methods for
stochastic-control problems in continuous time. Barles and Souganidis
\cite{barles1991convergence} give the monotone-stable-consistent convergence
criterion for fully nonlinear second-order equations. This paper follows that
route: the funding diffusion is approximated by a birth--death generator with
nonnegative transition rates, and the explicit update enforces a CFL condition.
The claim is therefore a finite-difference HJB policy with a standard
convergence path, not an analytical Riccati solution for the full arrival
Hamiltonian.

This distinction matters for the paper's positioning. Closed-form AS-type
formulas are elegant, but the funding term creates a two-dimensional
inventory--funding state. A projected quadratic ansatz is useful for intuition
and is retained in the appendix, yet the empirical strategy uses the full
discrete HJB table. The resulting contribution is computational and empirical:
it shows how to include stochastic funding in the quote-recovery loop and then
tests the resulting policy across assets.

\subsection{AMMs, DeFi liquidity, and LVR}

Automated market makers are the DeFi analogue of liquidity provision, but their
control variables differ sharply from an order-book market maker's quote
offsets. Hanson \cite{hanson2007logarithmic} introduced logarithmic market
scoring rules, an early constant-function-style mechanism for automated
liquidity in prediction markets. Uniswap v2 and v3 then made constant-product
and concentrated-liquidity AMMs central to DeFi market structure
\cite{adams2020uniswap,adams2021uniswap}. Angeris and Chitra
\cite{angeris2020improved} analyze constant-function market makers and price
oracles, while Angeris, Evans, and Chitra \cite{angeris2021replicating} connect
AMM payoffs to replicating portfolios. Angeris et al.
\cite{angeris2022optimal} study optimal routing across CFMMs, emphasizing that
AMM execution is constrained by the geometry of invariant curves rather than by
discrete quote placement.

The DeFi literature also documents frictions that are absent from a frictionless
HJB. Schar \cite{schar2021defi} gives a broad overview of smart-contract-based
financial markets. Werner et al. \cite{werner2022sok} survey DeFi protocol
risks and composability. Daian et al. \cite{daian2020flash} show that public
mempools and transaction ordering create miner/extractor value and frontrunning
risk. Mohan \cite{mohan2022amm} provides a primer on AMM mechanics for
financial applications. Aoyagi \cite{aoyagi2020liquidity} studies liquidity
provision by automated market makers as an economic problem, including
competition between informed and uninformed flow.

Loss-versus-rebalancing (LVR) provides the closest DeFi motivation for this
paper's empirical design. Milionis et al. \cite{milionis2022lvr} formalize the
loss borne by AMM liquidity providers relative to continuous rebalancing.
Cartea, Drissi, and Monga \cite{cartea2023predictable} study predictable loss
and optimal liquidity provision in decentralized finance, and their later work
\cite{cartea2025execution} connects AMM liquidity with execution and
speculation. This paper does not model a constant-product pool or a
concentrated-liquidity range. It borrows the DeFi question--how should a
liquidity supplier react to predictable state variables and adverse execution
channels--but places it in a perpetual DEX quote-control problem.

\subsection{Perpetual contracts and funding dynamics}

Perpetual contracts differ from fixed-maturity futures because the funding
mechanism keeps the contract price linked to the underlying index. The funding
rate is therefore not just a fee; it is a state variable that transfers cash
between longs and shorts and can shape the economics of inventory. Alexander et
al. \cite{alexander2020bitmex} study price discovery and informational
efficiency in Bitcoin derivatives markets, giving an empirical backdrop for
crypto derivatives microstructure. He et al. \cite{he2022fundamentals} study
fundamentals of perpetual futures, and Ackerer et al.
\cite{ackerer2023perpetual} analyze arbitrage and price formation in perpetual
markets. Kim and Park \cite{kim2025designing} study funding-rate design for
perpetual futures, emphasizing that funding-rule choices can materially affect
market quality.

The two closest references for the present paper are Kharat
\cite{kharat2025funding} and Donnelly, Lin, and Lorig
\cite{donnelly2026liquidation}. Kharat estimates stochastic funding-rate
dynamics with jumps, which is directly relevant to the calibration evidence in
Section~\ref{sec:calibration}. The same pattern appears here: a Gaussian OU
baseline is tractable and mean-reverting, but jump diagnostics are too strong
to ignore. Donnelly, Lin, and Lorig study optimal liquidation of perpetual
contracts with funding in the wealth dynamics. The difference is the control
surface. Liquidation controls how an existing inventory is reduced. Market
making controls the bid and ask quotes that create future inventory through
random fills.

The gap filled by the present paper is therefore precise. Existing
market-making models give quote recovery from inventory value differences, but
they generally omit perpetual funding. Existing AMM and LVR papers explain
DeFi liquidity losses, but they do not solve a two-sided perpetual quote-control
problem. Existing perpetual-control papers include funding, but they focus on
execution or liquidation rather than liquidity provision. This paper combines
these ingredients into a funding-aware finite-difference HJB and tests whether
the resulting quote policy adds value over funding-unaware AS benchmarks under
common execution assumptions.

\subsection{Novelty map and scope control}

The above comparison also clarifies what the paper should not claim. It should
not claim to dominate the AS literature in general, because the AS model solves
a different baseline problem and is intentionally parsimonious. The appropriate
claim is conditional: once a perpetual market maker observes a persistent
funding state, the continuation value of inventory depends on that state, and a
quote rule that ignores it leaves a measurable state variable unused. The HJB
does not replace the AS logic; it augments the AS inventory value difference
with a funding dimension. In the zero-funding limit, the quote recovery reduces
to the symmetric exponential-intensity spread and, with inventory penalties, to
the familiar inventory skew.

The paper should also avoid overstating the AMM connection. AMMs and order-book
market makers both provide liquidity, but their control surfaces are not the
same. A concentrated-liquidity LP chooses a price range and earns fees when
arbitrageurs or traders move through that range. The perpetual market maker
considered here chooses bid and ask offsets every decision period and faces
random fills. The common theme is state-dependent liquidity provision, not
identical mechanism design. This is why LVR and predictable-loss papers belong
in the motivation and discussion, while the mathematical core comes from
quote-control HJBs.

Finally, the perpetual-funding literature motivates a modeling compromise.
Funding data are mean reverting enough for a diffusion state to be useful, but
the residuals are not Gaussian in a literal sense. The empirical section
therefore treats Gaussian OU as the tractable control baseline and OU-plus-jump
as the econometric warning sign. A stronger future paper would put jumps
directly in the HJB through a nonlocal generator. This paper makes the smaller
claim that even the diffusion baseline is informative: if funding is a state
variable in the cash account, the market maker's marginal value of buying or
selling one more contract should depend on both inventory and funding.

\section{Mathematical Preliminaries}
\label{sec:prelim}

Work on a filtered probability space
$(\Omega,\mathcal{F},(\mathcal{F}_t)_{t\ge 0},\mathbb{P})$ satisfying the usual
conditions. The filtration contains the funding Brownian motion and the
controlled bid/ask order-arrival point processes. The mark price is observed
exogenously in the empirical backtest. The control equation used below is
therefore written for the reduced state $(t,X_t,q_t,f_t)$, where $X_t$ is cash,
$q_t$ is perpetual inventory, and $f_t$ is the cash funding amount per unit
inventory.

\begin{assumption}[Admissible controls and compact state]
Bid and ask offsets $\delta^b_t,\delta^a_t$ are predictable processes taking
values in a compact interval $[\delta_{\min},\delta_{\max}]$. Inventory is
constrained to a finite grid
\[
\mathcal{Q}=\{q_{\min},q_{\min}+\Delta q,\ldots,q_{\max}\}.
\]
Orders that would move inventory outside this grid are not submitted. The
funding diffusion is solved on a truncated interval
$[f_{\min},f_{\max}]$. Fill intensities are bounded on the admissible quote
set. Terminal and running inventory penalties are nonnegative.
\end{assumption}

These restrictions are not merely numerical conveniences. They make the
control problem finite on the computational domain, give compact maximization
sets for the Hamiltonian, and match the backtester, which blocks fills that
would breach the inventory limit. They also make the quote-recovery map
measurable: for each state, the optimal bid and ask offsets are deterministic
functions of the local value differences.

The dynamic-programming principle states that the value function at time $t$
equals the supremum of the expected immediate reward plus the continuation
value after a small time step. Formally proving the principle for controlled
diffusions with point-process jumps requires standard regularity and
measurability conditions. The standard stochastic-control route is used: derive
the HJB from the dynamic-programming principle, recover measurable maximizers
from the Hamiltonian, and verify that a sufficiently smooth solution is the
value function. References for the continuous-time stochastic-control background are
Pham \cite{pham2009continuous} and Touzi \cite{touzi2013optimal}.

\begin{theorem}[Verification route]
Suppose a function $u$ is continuous on the compact state space, $C^1$ in time
and $C^2$ in the funding coordinate in the interior, satisfies the terminal
condition and the HJB equation, and has predictable measurable Hamiltonian
maximizers. Then $u$ coincides with the value function on the truncated domain,
and the associated feedback controls are optimal among admissible controls.
\end{theorem}

The proof is standard. Apply Ito's formula with jumps to the candidate value
process, use the HJB inequality for arbitrary admissible controls to obtain an
upper bound, and use the maximizing selector to attain equality. For an
analogous diffusion verification theorem, see Pham
\cite[Theorem~3.5.2]{pham2009continuous}; the inventory jump component is
handled through the finite-difference point-process bookkeeping used in CJP
\cite[Section~6.4]{cartea2015algorithmic}. The inventory state is discrete, so
the jump part appears through finite differences in $q$ rather than through
derivatives. The funding state is continuous in the limiting model and
contributes drift and diffusion terms.

The numerical policy used empirically is a finite-difference approximation, so
the relevant solution concept is viscosity convergence rather than classical
smooth verification. The Barles--Souganidis theorem gives convergence of
monotone, stable, consistent schemes for fully nonlinear second-order equations
when the limiting equation has a comparison principle
\cite{barles1991convergence}. Section~\ref{sec:hjb} constructs the funding
generator as a birth--death chain with nonnegative transition rates and
enforces an explicit CFL condition. On the truncated compact domain, this gives
a monotone approximation of the diffusion HJB. The convergence statement in
this paper is therefore conditional on the standard comparison principle for
the limiting HJB.

\begin{remark}
The convergence statement applies to the diffusion HJB baseline. The empirical
calibration in Section~\ref{sec:calibration} shows that OU-plus-jump funding
fits better than Gaussian OU. Adding jumps to the control equation would add a
nonlocal generator term and would require a separate monotone discretization.
That extension is left outside the present paper.
\end{remark}

\section{Model Setup}
\label{sec:model}

\subsection{State variables and units}

The full economic state is $(t,X_t,q_t,S_t,f_t)$. Cash $X_t$ is measured in
quote currency. Inventory $q_t$ is the number of perpetual contracts held by
the market maker, with positive $q_t$ denoting a long position. The mark price
$S_t$ is the current mid or mark price of the perpetual contract. The funding
state $f_t$ is a cash amount per unit inventory per hour. Hyperliquid reports a
fractional funding rate $F_t$; the empirical implementation uses
\[
  f_t \approx S_tF_t.
\]
This cash scaling is important because the HJB compares funding payments with
spread capture, both of which are cash quantities.

The sign convention is that the cash account pays $q_tf_t\,dt$. Thus a long
position pays funding when $f_t>0$ and receives funding when $f_t<0$. A short
position has the opposite exposure. This sign convention is consistent with the
cash equation below and with the backtest accounting.

\subsection{Dimensional consistency}

The empirical funding feed is quoted as a fractional rate per funding period,
while quote offsets and cash PnL are measured in quote currency. This
difference is easy to miss and is important for the HJB. Let $F_t$ denote the
fractional funding rate per hour and let $S_t$ be the current mark price. For
one unit of inventory, the hourly cash transfer is approximately
\[
  S_tF_t.
\]
The reduced HJB therefore uses
\[
  f_t=S_tF_t
\]
as the funding state. With time measured in hours, $q_tf_t\,dt$ has cash units:
contracts times cash per contract per hour times hours. This puts the funding
source term $-qf$ in the same units as spread capture from a fill.

In the historical simulator, funding is applied at the observed hourly
timestamps rather than continuously. If $\tau_m$ is a funding timestamp and
$\Delta\tau=1$ hour, the discrete cash update is
\[
  X_{\tau_m^+}
  =
  X_{\tau_m^-}
  -
  q_{\tau_m^-}S_{\tau_m}F_{\tau_m}\Delta\tau .
\]
The continuous equation in the HJB is the small-step analogue of this update.
Using cash-scaled funding is also the reason the policy can compare a
one-basis-point quote change with a funding signal without an arbitrary
multiplier.

The quote offsets $\delta^a$ and $\delta^b$ are cash distances from the mid
price per unit contract. If the inventory grid step is $\Delta q$, then an ask
fill earns $\delta^a\Delta q$ in spread revenue after the mark-to-market term
cancels, and a bid fill earns $\delta^b\Delta q$. The implementation fixes the
quote size and absorbs it into the value-difference convention used by quote
recovery. This is why Section~\ref{sec:hjb} writes the first-order condition
per quote unit. The backtester and HJB use the same sign convention: positive
inventory means long, bid fills increase inventory, ask fills decrease it, and
positive funding is a cash debit to long inventory.

\subsection{Funding dynamics}

The diffusion baseline models the cash-scaled funding state as an
Ornstein--Uhlenbeck process,
\[
  df_t=\kappa(\bar f-f_t)\,dt+\sigma_f\,dW^f_t,
\]
where $\kappa>0$ is the mean-reversion speed, $\bar f$ is the long-run funding
level, and $\sigma_f$ is the funding volatility. Time is measured in hours in
the calibration and in the numerical HJB. The OU process is chosen because its
transition density is Gaussian and closed form, making calibration and grid
construction transparent. It is not claimed to be the final econometric model;
Section~\ref{sec:calibration} shows that jumps materially improve the funding
likelihood.

The mark price is replayed from historical data in the backtest. One could
write a joint diffusion for $(S_t,f_t)$ with correlated Brownian motions, but
the final numerical policy does not use a predictive price drift. This is a
deliberate modeling choice: the experiment is intended to isolate the value of
funding-aware inventory control rather than mix the result with directional
price forecasting. Price risk still enters through mark-to-market accounting
and through the terminal liquidation value.

\subsection{Order arrivals and inventory}

The market maker chooses bid and ask distances
$\delta^b_t,\delta^a_t\ge \delta_{\min}$ from the mid price. When a bid order
fills, inventory increases by the quote size $\Delta q$; when an ask order
fills, inventory decreases by $\Delta q$. The inventory dynamics are
\[
  dq_t=\Delta q\,dN^b_t-\Delta q\,dN^a_t.
\]
The fill processes have controlled intensities
\[
  \lambda^{b/a}(\delta)=\Lambda e^{-k\delta},
\]
where $\Lambda$ is the effective at-touch fill intensity and $k$ controls how
quickly fill probability decays as quotes move away from the mid. In the
backtest, $\Lambda$ and $k$ are calibrated from official Hyperliquid crossed
fills joined to the local one-minute L2 panel. The calibration is an effective
fill model rather than a queue-priority model.

Inventory is constrained to $[q_{\min},q_{\max}]$. If a bid fill would push
inventory above $q_{\max}$, the bid side is blocked; if an ask fill would push
inventory below $q_{\min}$, the ask side is blocked. This hard constraint is
used both in the HJB grid and in the simulator.

\subsection{Cash dynamics and objective}

When an ask fills, the market maker sells at $S_t+\delta^a_t$; when a bid
fills, the market maker buys at $S_t-\delta^b_t$. Funding is paid continuously
in the reduced model and hourly in the backtest. The cash dynamics are
\[
  dX_t=(S_t+\delta^a_t)\Delta q\,dN^a_t
       -(S_t-\delta^b_t)\Delta q\,dN^b_t
       -q_tf_t\,dt.
\]
The first two terms are spread capture from executed quotes. The last term is
the funding payment on the running inventory.

The objective is linear terminal wealth with running and terminal inventory
penalties:
\[
v(t,x,q,s,f)=
\sup_{\delta^a,\delta^b}
\mathbb{E}_{t,x,q,s,f}\left[
X_T+q_TS_T-\alpha q_T^2
-\phi\int_t^T q_u^2\,du
\right].
\]
The terminal penalty $\alpha q_T^2$ discourages ending the horizon with a large
position, while the running penalty $\phi\int q_u^2du$ discourages large
inventory throughout the horizon. This linear-utility specification is not a
complete risk preference model, but it is standard for numerical market-making
baselines and keeps the funding mechanism visible in the HJB. The benchmark AS
policy is recovered by shutting down the funding state and using the
funding-unaware inventory-risk quote rule.

\section{Finite-Difference HJB and Quote Recovery}
\label{sec:hjb}

This section derives the control equation used by the final empirical
strategy. Earlier projected Riccati approximations were useful diagnostics, but
they did not provide a robust closed-form treatment of the full arrival
Hamiltonian. The final policy therefore solves the reduced inventory--funding
HJB numerically and recovers quotes from the value table.

\subsection{Reduction by mark-to-market wealth}

Because the objective is linear in terminal cash and mark-to-market inventory
value, use the ansatz
\[
  v(t,x,q,s,f)=x+qs+\theta(t,q,f).
\]
The function $\theta$ is the continuation value after removing current cash
and current mark-to-market inventory value. It captures the future value of
inventory, funding exposure, running penalties, terminal penalties, and fill
opportunities. Under the no-price-drift policy used in the final experiments,
the price state does not appear in $\theta$. Price still enters realized PnL in
the simulator, but it is not used to forecast future returns.

To see the reduction, consider an ask fill. Inventory changes from $q$ to
$q-\Delta q$ and cash increases by $(s+\delta^a)\Delta q$. The change in
$x+qs$ from this fill is
\[
  (s+\delta^a)\Delta q + (q-\Delta q)s-qs
  =\delta^a\Delta q.
\]
The mid-price component cancels. Thus the ask-fill contribution to the reduced
value is spread capture plus the change in $\theta$. A bid fill analogously
contributes $\delta^b\Delta q+\theta(t,q+\Delta q,f)-\theta(t,q,f)$.

\subsection{Reduced HJB}

The reduced value function solves
\[
\begin{aligned}
0={}&\partial_t\theta
 +\kappa(\bar f-f)\partial_f\theta
 +\frac{1}{2}\sigma_f^2\partial_{ff}\theta
 -qf-\phi q^2 \\
&+\mathbf{1}_{q>q_{\min}}H^a(t,q,f)
 +\mathbf{1}_{q<q_{\max}}H^b(t,q,f),
\end{aligned}
\]
with terminal condition
\[
  \theta(T,q,f)=-\alpha q^2.
\]
The source term $-qf$ is the structural difference from classical
Avellaneda--Stoikov. It couples inventory and funding: when $q$ and $f$ have
the same sign, funding is costly; when they have opposite signs, funding is
beneficial. This is the mechanism that makes the quote skew funding-aware.

For quote size $\Delta q$, the ask and bid Hamiltonians are
\[
H^a(t,q,f)=
\sup_{\delta^a\ge\delta_{\min}}
\Lambda e^{-k\delta^a}
\left[
\delta^a\Delta q+\theta(t,q-\Delta q,f)-\theta(t,q,f)
\right],
\]
and
\[
H^b(t,q,f)=
\sup_{\delta^b\ge\delta_{\min}}
\Lambda e^{-k\delta^b}
\left[
\delta^b\Delta q+\theta(t,q+\Delta q,f)-\theta(t,q,f)
\right].
\]
In the implementation, $\Delta q$ is absorbed into the inventory grid and quote
size conventions, so the displayed quote recovery below is written per quote
unit. The economic interpretation is unchanged: each side balances marginal
spread revenue against the marginal continuation value of changing inventory.

\subsection{Pointwise quote recovery}

Let
\[
  A(t,q,f)=\theta(t,q-\Delta q,f)-\theta(t,q,f),
  \qquad
  B(t,q,f)=\theta(t,q+\Delta q,f)-\theta(t,q,f).
\]
Ignoring the quote floor for a moment, the ask Hamiltonian maximizes
$e^{-k\delta}(\delta+A)$ and the bid Hamiltonian maximizes
$e^{-k\delta}(\delta+B)$. The first-order condition is
\[
  1-k(\delta+A)=0
\]
for the ask side and analogously for the bid side. Hence
\[
\delta^{a*}(t,q,f)=\frac{1}{k}-A(t,q,f),
\qquad
\delta^{b*}(t,q,f)=\frac{1}{k}-B(t,q,f).
\]
The implemented policy applies
\[
  \delta^{a,b*}\leftarrow \max(\delta_{\min},\delta^{a,b*}).
\]
At the inventory boundaries, the side that would breach the inventory limit is
removed from the Hamiltonian and is represented in the simulator by a very
large quote distance.

The formula has the expected sign behavior. If buying one more unit has low
continuation value, then $B$ is negative and the bid quote widens. If selling
one unit reduces a costly long inventory, then $A$ is positive and the ask quote
tightens. Funding changes these value differences because it changes the
future cash-flow value of holding positive or negative inventory.

\subsection{Monotone finite-difference scheme}

The solver uses a tensor grid
\[
  t_i\in[0,T],\qquad q_j=q_{\min}+j\Delta q,\qquad
  f_\ell\in[f_{\min},f_{\max}].
\]
The terminal slice is initialized by
\[
  \theta_{N,j,\ell}=-\alpha q_j^2.
\]
Backward in time, the explicit update is
\[
\theta_{i,j,\ell}
=\theta_{i+1,j,\ell}
+\Delta t\left(
\mathcal{L}^f_h\theta_{i+1,j,\ell}
-q_jf_\ell-\phi q_j^2
+H^a_{i+1,j,\ell}+H^b_{i+1,j,\ell}
\right).
\]

The funding generator is discretized as a birth--death chain. Let
$b(f)=\kappa(\bar f-f)$ and let $\Delta f$ be the funding-grid spacing. In the
interior, define nonnegative rates
\[
r^+_\ell=\frac{\sigma_f^2}{2\Delta f^2}+\frac{b(f_\ell)^+}{\Delta f},
\qquad
r^-_\ell=\frac{\sigma_f^2}{2\Delta f^2}+\frac{b(f_\ell)^-}{\Delta f}.
\]
Then
\[
\mathcal{L}^f_h\theta_{j,\ell}
=r^+_\ell(\theta_{j,\ell+1}-\theta_{j,\ell})
+r^-_\ell(\theta_{j,\ell-1}-\theta_{j,\ell}).
\]
At $f_{\min}$ and $f_{\max}$, transitions that would leave the truncated domain
are suppressed. This is an upwind Markov-chain approximation of the diffusion
generator. Its off-diagonal coefficients are nonnegative, which is the key
property needed for monotonicity.

Let $\lambda_{\max}=\Lambda e^{-k\delta_{\min}}$. Since at most two quote sides
can be active, the explicit update is monotone when
\[
\Delta t\left(
\max_\ell(r^+_\ell+r^-_\ell)+2\lambda_{\max}
\right)\le 1.
\]
The code computes this CFL bound and rejects grids that violate it. The final
selected cross-asset configurations use $N_t=2048$, which satisfies the bound
under both official-fill calibrations used in Section~\ref{sec:backtest}.

\subsection{Classical AS limit}

The reduced equation nests the funding-unaware AS logic. If $f\equiv 0$,
$\bar f=0$, $\sigma_f=0$, and inventory penalties are shut down, then
$\theta$ is constant in $q$ and $f$. The quote recovery formulas give
\[
  \delta^{a*}=\delta^{b*}=\frac{1}{k},
\]
the symmetric risk-neutral exponential-intensity spread. With inventory
penalties but no funding, the value differences depend only on inventory and
generate the usual AS inventory skew. Funding adds a second state dimension:
the inventory value difference now depends on whether the current funding state
makes holding inventory attractive or costly.

\subsection{What is not claimed}

The result is a numerical HJB policy, not a closed-form Riccati solution. The
linear-quadratic Riccati prototype in the appendix is useful for intuition, but
the full arrival Hamiltonian is handled numerically in the reported strategy.
The result is also a diffusion HJB. The OU-plus-jump evidence in
Section~\ref{sec:calibration} motivates a nonlocal jump-HJB extension but does
not enter the final backtest table.

\section{Empirical Calibration}
\label{sec:calibration}

\subsection{Data alignment}

The empirical pipeline uses Hyperliquid hourly funding observations and a local
one-minute Hyperliquid L2 panel for ETH, BTC, and SOL. Funding is observed as a
fractional rate $F_t$. For the HJB, the state is converted to a cash funding
amount by multiplying by a representative mid price when constructing the grid,
while policy evaluation uses the contemporaneous signal $S_tF_t$. The backtest
holdout window is 26 November 2025 through 31 December 2025. Fill intensities
are calibrated on earlier official-fill data so the holdout is not used to tune
the execution model.

Notation is fixed throughout the paper as follows. Uppercase $F_t$ denotes the
observed fractional funding rate, while lowercase $f_t$ denotes the cash-scaled
state used in the HJB:
\[
  f_t = S_t F_t .
\]
The likelihoods in this section are estimated on $F_t$ because that is the
exchange-reported funding series. The HJB and backtest then apply the cash
conversion before quote recovery or funding accounting. Thus every occurrence
of lowercase $f$ in the control equation is cash-scaled.

The funding process is calibrated first, because it determines whether the
funding state is worth including in the HJB. The pre-flight criterion is
mean-reversion over economically relevant horizons. If funding behaved like a
unit-root process over the sample, an OU control state would be the wrong
baseline. In the observed ETH/BTC/SOL samples, half-lives are measured in hours
rather than weeks, so funding is persistent enough to matter but not so slow
that the state is effectively constant.

\subsection{OU transition likelihood}

For hourly observations $(F_{t_i})$, the Gaussian OU model has the exact
transition
\[
F_{t+\Delta}
\mid F_t
\sim
\mathcal{N}\left(
\theta+(F_t-\theta)e^{-\kappa\Delta},
\frac{\sigma^2}{2\kappa}(1-e^{-2\kappa\Delta})
\right).
\]
The parameters $(\kappa,\theta,\sigma)$ are estimated by maximizing the sum of
log transition densities on the training sample. The reported half-life is
\[
  t_{1/2}=\frac{\log 2}{\kappa}.
\]
Train/test splits are used to ensure that the fitted diffusion is not only a
within-sample description. The standardized residuals are then checked for
skewness, excess kurtosis, and normality failures. These residual diagnostics
are important because the HJB uses the diffusion as a tractable baseline even
when the funding data are not Gaussian.

Table~\ref{tab:calibration-summary} reports the main cross-asset diagnostics.
The half-lives range from roughly two to six hours. Funding-price innovation
correlations are small, so the final HJB does not rely on a predictive
price-funding cross term. The OU-plus-jump likelihood gains are large for all
three assets, which is evidence against treating Gaussian OU as a complete
funding model.

\begin{table}[h]
\centering
\small
\begin{tabular}{lrrrr}
\toprule
Asset & OU half-life (hours) & Jump prob./hour & OU+jump LL gain & Funding-price $\rho$ \\
\midrule
ETH & 5.560 & 2.05\% & 2816.34 & 0.0026 \\
BTC & 4.071 & 1.24\% & 4417.03 & 0.0278 \\
SOL & 2.310 & 0.16\% & 7149.78 & -0.0151 \\
\bottomrule
\end{tabular}
\caption{Funding-rate calibration summary. Half-lives and likelihoods are
estimated on the observed fractional funding rate $F_t$; the HJB state is the
cash-scaled quantity $f_t=S_tF_t$. Jump probabilities are the fitted
Bernoulli-normal arrival probabilities per hour. The OU+jump likelihood gain
is the increase in train log-likelihood over the Gaussian OU transition model.}
\label{tab:calibration-summary}
\end{table}

\begin{figure}[h]
\centering
\includegraphics[width=\textwidth]{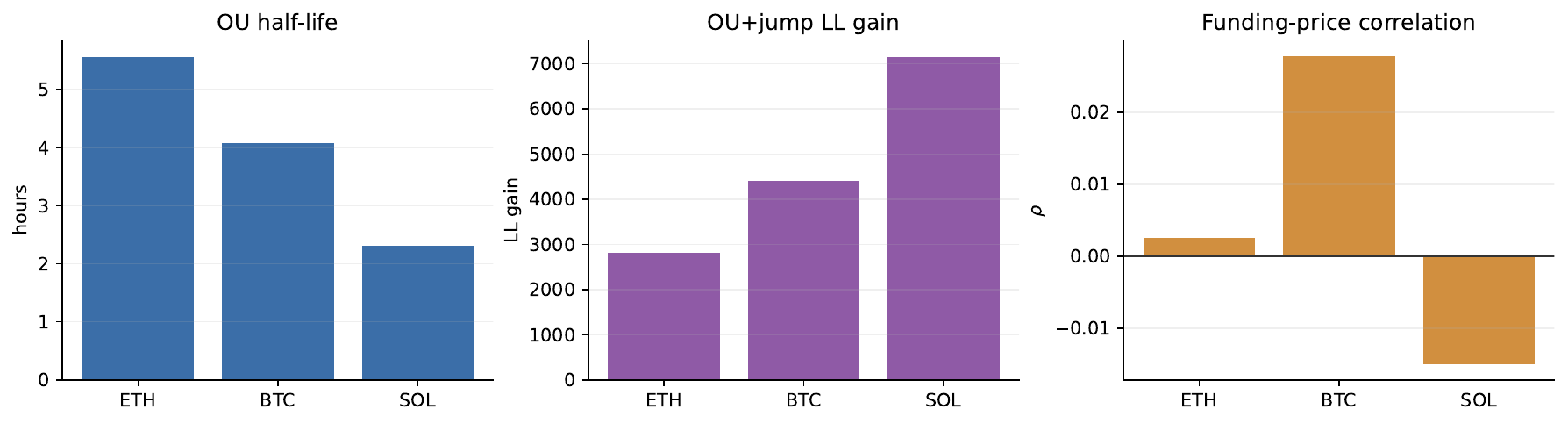}
\caption{Cross-asset funding diagnostics used by the HJB calibration. Funding
mean-reversion is fast enough to be a useful state variable, while the
OU-plus-jump likelihood gains show that Gaussian OU is only a tractable
baseline.}
\label{fig:calibration-summary}
\end{figure}

\subsection{OU-plus-jump diagnostic}

The jump diagnostic uses a Bernoulli-normal transition mixture on the observed
fractional funding series $F_t$. Let
\[
  m_i=\theta+(F_{t_i}-\theta)e^{-\kappa\Delta_i},
  \qquad
  v_i=\frac{\sigma^2}{2\kappa}(1-e^{-2\kappa\Delta_i}),
\]
and let $Z_i\sim\mathrm{Bernoulli}(p_i)$ with
\[
  p_i=1-e^{-\lambda_J\Delta_i}.
\]
The transition approximation is
\[
  F_{t_{i+1}}
  =
  m_i+\varepsilon_i+Z_iJ_i^F,
  \qquad
  \varepsilon_i\sim\mathcal{N}(0,v_i),
  \qquad
  J_i^F\sim\mathcal{N}(\mu_J,\sigma_J^2),
\]
with independent diffusion and jump shocks. This is a one-jump small-time
approximation rather than a full compound-Poisson expansion. The fitted hourly
jump probabilities are $2.05\%$ for ETH, $1.24\%$ for BTC, and $0.16\%$ for
SOL. This is not yet embedded in the control equation, but it is a useful
econometric stress test for the diffusion assumption. The improvement in log
likelihood is largest for SOL and material for ETH and BTC as well.

The implication is methodological. The diffusion HJB should be presented as a
baseline because it is tractable, monotone-discretizable, and already captures
mean reversion. It should not be presented as the final law of funding-rate
dynamics. A future jump-HJB would convert fractional jumps into cash jumps
$J^f\approx S_tJ^F$ and add a nonlocal term of the form
\[
  \lambda_J
  \mathbb{E}\left[
    \theta(t,q,f+J^f)-\theta(t,q,f)
  \right],
\]
where $J^f$ is the cash-scaled jump size. That term would change the value of carrying
inventory in jump-prone funding regimes and may materially affect quote skew.

\subsection{Fill-intensity calibration}

The execution model is calibrated from official Hyperliquid crossed fills
joined to the one-minute L2 panel. For each asset and quote-distance bucket,
the pipeline estimates an effective fill curve of the form
\[
  \lambda(\delta)=\Lambda e^{-k\delta}.
\]
The baseline calibration is \texttt{volume\_minute}: a minute counts as a hit
at distance $d$ if cumulative crossed volume at that threshold is at least the
strategy quote size. This is stricter than counting every trade and avoids the
unrealistically high intensities obtained from raw fill counts. The alternative
robustness calibration is \texttt{minute\_hit}: a minute counts as a hit if the
threshold is touched at least once, regardless of whether volume reaches the
quote size.

Both calibrations are proxy execution models. They do not model queue position,
latency, maker priority, or adverse selection conditional on being filled.
Their role in this paper is narrower: they provide two official-fill-based
intensity curves under which the same policies can be compared. A result that
only survives one proxy fill curve would be fragile; the ETH/BTC HJB result in
Section~\ref{sec:backtest} survives both.

\section{Backtest}
\label{sec:backtest}

The backtest replays the one-minute Hyperliquid L2 mid-price panel and hourly
funding observations. At each minute the strategy posts bid and ask quotes,
fills are simulated under an exponential distance-to-mid intensity model, and
funding payments are applied hourly to the running perpetual inventory. The
arrival model is calibrated from official Hyperliquid crossed fills using a
quote-size-aware minute-volume proxy. This is not a queue-position or
latency-aware execution model, so the empirical results should be read as a
controlled comparison of policies under a common fill proxy.

The simulation is event driven at the one-minute frequency. At each timestamp,
the policy observes elapsed time, current inventory, the latest fractional
funding observation $F_t$, and the mid price. HJB policies convert this to the
cash-scaled signal $f_t=S_tF_t$ before quote recovery. The policy returns bid
and ask distances. Conditional on those
distances, bid and ask fills are sampled independently using the calibrated
intensity curve and the elapsed time since the previous observation. If a fill
would breach the inventory limit, that side is blocked. Funding payments are
applied when the replay crosses an hourly boundary. Final equity is cash plus
mark-to-market perpetual inventory, with optional hedge accounting disabled in
the final table.

The reported metrics are chosen to separate profitability from inventory risk.
Final equity is the terminal cash plus mark-to-market inventory value. The
confidence interval is $1.96$ times the standard error across seeds. Win rate
is paired by seed against \texttt{pure\_as}; this removes simulation-noise
differences caused by comparing different random seeds. Inventory RMS is the
root mean square of inventory over the path and is the main risk-exposure
metric. Max drawdown is computed from the simulated equity path. Fill rate is
the realized fraction of quoted sides that fill.

The final robustness run freezes the validation-selected parameters and uses
one hundred simulation seeds, $1,\ldots,100$, on the same holdout window:
26 November 2025 through 31 December 2025. The final table includes four
policies:
\begin{itemize}
    \item classical Avellaneda--Stoikov without funding awareness
    (\texttt{pure\_as});
    \item a risk-scaled AS diagnostic with quote size and inventory limit scaled
    to approximately match the HJB inventory RMS (\texttt{pure\_as\_scaled});
    \item the selected finite-difference funding-aware HJB policy
    (\texttt{hjb\_fd});
    \item a risk-calibrated funding-aware quote rule used as a practical
    benchmark (\texttt{risk\_calibrated}).
\end{itemize}
The carry-overlay and projected Riccati variants are excluded from the final
table because they are not robust across assets.

The scaled AS row is a diagnostic rather than a production benchmark: it changes
quote size and inventory capacity while keeping the same effective-fill curve.
Its purpose is to test whether apparent gains are merely compensation for more
inventory risk.

The selected HJB parameters are fixed before the final one-hundred-seed run:
\[
\begin{array}{c|cc}
\text{Asset} & \alpha & \phi \\
\hline
\text{ETH} & 5\times 10^{-4} & 10^{-4} \\
\text{BTC} & 2\times 10^{-3} & 10^{-4} \\
\text{SOL} & 10^{-3} & 5\times 10^{-5}.
\end{array}
\]

Table~\ref{tab:final-backtest} reports means over the one hundred seeds. The confidence
interval is $1.96$ times the standard error of final equity. Win rate is the
paired fraction of seeds in which the policy's final equity exceeds
\texttt{pure\_as} for the same seed.

\begin{table}[h]
\centering
\small
\begin{tabular}{llrrrrr}
\toprule
Asset & Policy & Final equity & 95\% CI & $\Delta$ vs AS & Win rate & Inv. RMS \\
\midrule
ETH & \texttt{pure\_as} & 74471.93 & 714.76 & 0.00 & -- & 5.7362 \\
ETH & \texttt{pure\_as\_scaled} & 46775.85 & 431.86 & -27696.09 & 0.00 & 3.6373 \\
ETH & \texttt{hjb\_fd} & 75784.42 & 459.81 & 1312.49 & 0.62 & 3.6475 \\
ETH & \texttt{risk\_calibrated} & 75492.43 & 345.24 & 1020.49 & 0.63 & 5.2318 \\
\midrule
BTC & \texttt{pure\_as} & 46808.28 & 711.09 & 0.00 & -- & 0.2974 \\
BTC & \texttt{pure\_as\_scaled} & 28832.62 & 412.79 & -17975.66 & 0.00 & 0.1832 \\
BTC & \texttt{hjb\_fd} & 47617.37 & 430.62 & 809.09 & 0.57 & 0.1831 \\
BTC & \texttt{risk\_calibrated} & 46737.00 & 688.79 & -71.28 & 0.52 & 0.2942 \\
\midrule
SOL & \texttt{pure\_as} & 85989.45 & 146.85 & 0.00 & -- & 20.0173 \\
SOL & \texttt{pure\_as\_scaled} & 196201.76 & 349.36 & 110212.31 & 1.00 & 43.2371 \\
SOL & \texttt{hjb\_fd} & 104860.37 & 279.55 & 18870.92 & 1.00 & 51.8427 \\
SOL & \texttt{risk\_calibrated} & 86068.29 & 144.22 & 78.83 & 0.63 & 20.1691 \\
\bottomrule
\end{tabular}
\caption{Final one-hundred-seed holdout results for selected policies. Final equity,
confidence intervals, and deltas are in quote-currency units.}
\label{tab:final-backtest}
\end{table}

\begin{figure}[h]
\centering
\includegraphics[width=\textwidth]{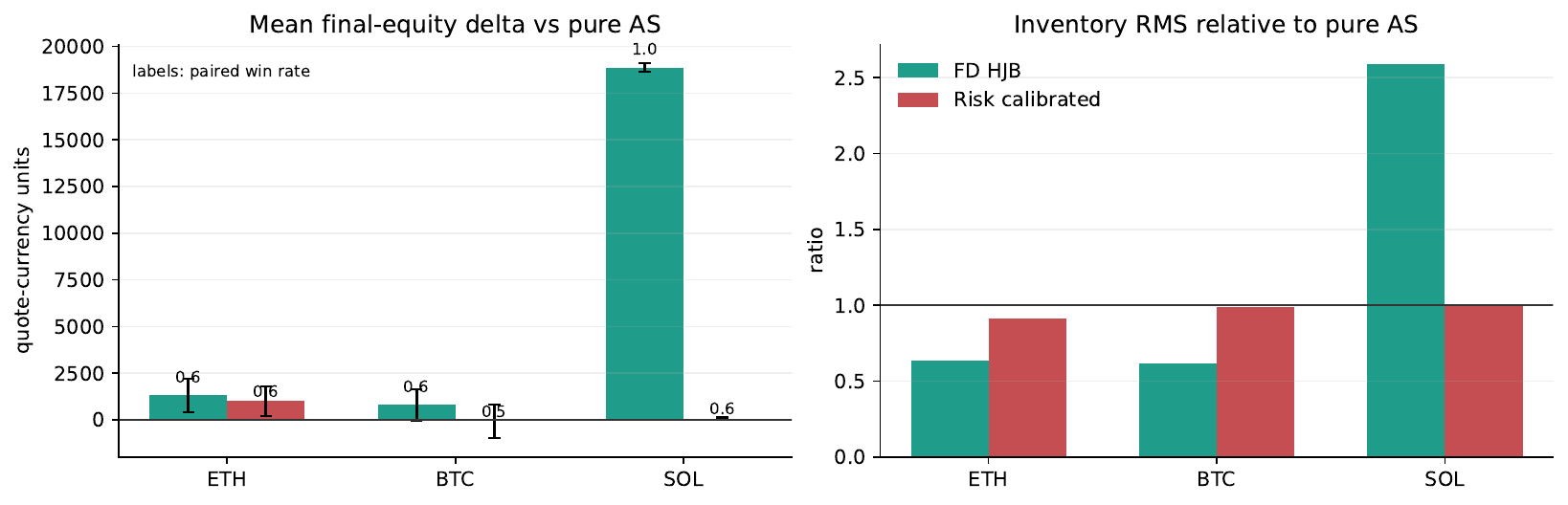}
\caption{Final one-hundred-seed holdout comparison. The left panel reports paired
final-equity deltas relative to \texttt{pure\_as}; labels above bars are paired
win rates. The right panel reports inventory RMS relative to \texttt{pure\_as}.}
\label{fig:final-backtest-summary}
\end{figure}

\begin{figure}[h]
\centering
\includegraphics[width=0.72\textwidth]{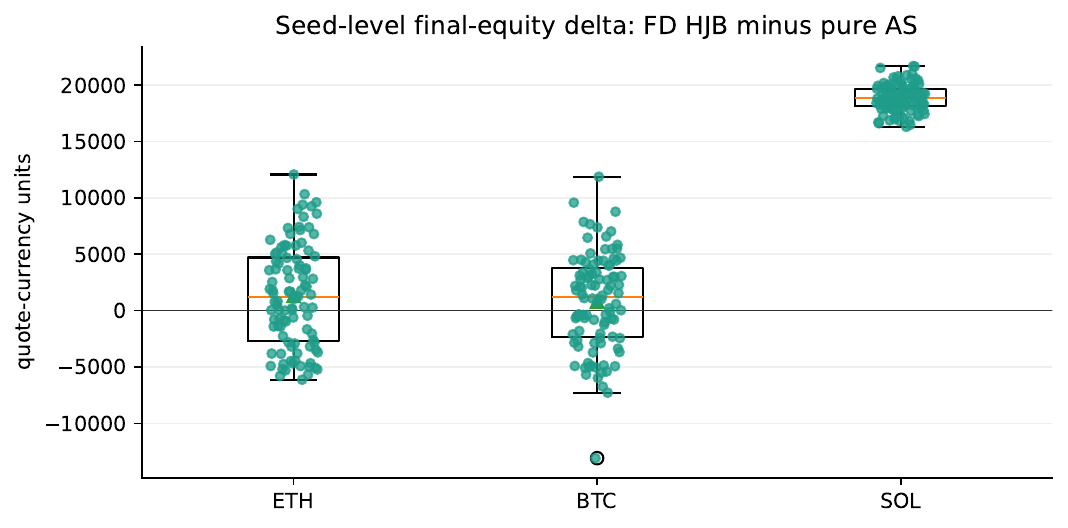}
\caption{Seed-level final-equity delta of the finite-difference HJB policy
relative to \texttt{pure\_as}. BTC is the weakest asset because the mean
improvement is not pathwise robust across simulation seeds.}
\label{fig:final-backtest-seeds}
\end{figure}

The main positive result is that the finite-difference HJB has higher mean
final equity than \texttt{pure\_as} in all three assets and substantially lower
inventory RMS in ETH and BTC. The 100-seed run also weakens two earlier
overstatements. First, BTC is no longer a clear negative pathwise result, but it
remains statistically modest: the HJB win rate is $0.57$ and the paired equity
delta has a wide confidence interval. Second, the SOL gain is not a Pareto
improvement. A risk-scaled AS diagnostic earns substantially more than the HJB
on SOL, so the SOL result should be read as the HJB taking a different
risk-return position rather than discovering free funding alpha.

The defensible claim is narrower. Under the calibrated effective-fill simulator,
the selected HJB improves average ETH/BTC performance while lowering inventory
RMS. The SOL gain is mainly a leverage/risk-appetite effect. Simple projected
funding-aware quote rules are not robust.

To check dependence on the fill proxy, the final experiment is repeated with a
second official-fill calibration. The baseline \texttt{volume\_minute} model
records whether crossed volume at a distance threshold is large enough to fill
the strategy's quote size. The alternative \texttt{minute\_hit} model records
only whether a threshold was touched at least once during the minute. This
raises effective fill rates but keeps the same train/holdout split, seeds, and
selected HJB penalties. Under \texttt{minute\_hit}, the HJB deltas versus
\texttt{pure\_as} are $1413.74$ on ETH, $1138.69$ on BTC, and $23452.57$ on
SOL, with paired win rates $0.59$, $0.63$, and $1.00$, respectively. ETH and
BTC inventory RMS remain lower than \texttt{pure\_as}; SOL is again dominated
by risk-scaled AS. Thus the ETH/BTC conclusion survives an alternative
official-fill calibration, while the execution model remains a proxy rather
than a queue-position or latency model.

\begin{table}[h]
\centering
\small
\begin{tabular}{llrrrr}
\toprule
Asset & Fill model & HJB $\Delta$ vs AS & Win rate & HJB Inv. RMS & AS Inv. RMS \\
\midrule
ETH & \texttt{volume\_minute} & 1312.49 & 0.62 & 3.6475 & 5.7362 \\
ETH & \texttt{minute\_hit} & 1413.74 & 0.59 & 3.5772 & 5.6865 \\
\midrule
BTC & \texttt{volume\_minute} & 809.09 & 0.57 & 0.1831 & 0.2974 \\
BTC & \texttt{minute\_hit} & 1138.69 & 0.63 & 0.1790 & 0.2954 \\
\midrule
SOL & \texttt{volume\_minute} & 18870.92 & 1.00 & 51.8427 & 20.0173 \\
SOL & \texttt{minute\_hit} & 23452.57 & 1.00 & 52.7873 & 18.6680 \\
\bottomrule
\end{tabular}
\caption{Alternative fill-model robustness. The HJB policy uses the same
selected penalties and the same seeds. ETH/BTC retain lower inventory RMS under
both official-fill proxy calibrations; SOL does not.}
\label{tab:fill-robustness}
\end{table}

As an additional robustness check, the same selected policies are evaluated on
automatically chosen three-day stress windows within the holdout panel. For
each asset, the windows correspond to the highest average funding, lowest
average funding, highest realized volatility, and a calm low-volatility/low-
funding window. Table~\ref{tab:stress-backtest} reports the finite-difference
HJB result relative to \texttt{pure\_as}. This diagnostic uses ten seeds due to
compute budget; the one-hundred-seed robustness evidence is restricted to the
full holdout window, and a one-hundred-seed stress-window rerun is left for a
future version.

\begin{table}[h]
\centering
\small
\begin{tabular}{llrrr}
\toprule
Asset & Window & $\Delta$ vs AS & Win rate & Inv. RMS ratio \\
\midrule
ETH & high funding & -305.35 & 0.40 & 0.63 \\
ETH & low funding & 126.08 & 0.70 & 0.63 \\
ETH & high volatility & -266.52 & 0.40 & 0.62 \\
ETH & calm & 371.84 & 0.80 & 0.62 \\
\midrule
BTC & high funding & 455.01 & 0.80 & 0.62 \\
BTC & low funding & 151.02 & 0.60 & 0.62 \\
BTC & high volatility & -134.22 & 0.50 & 0.61 \\
BTC & calm & -75.99 & 0.40 & 0.61 \\
\midrule
SOL & high funding & 1615.99 & 1.00 & 2.56 \\
SOL & low funding & 1370.84 & 1.00 & 2.60 \\
SOL & high volatility & 1673.77 & 1.00 & 2.54 \\
SOL & calm & 1632.51 & 1.00 & 2.49 \\
\bottomrule
\end{tabular}
\caption{Stress-window HJB robustness over ten seeds. The inventory RMS ratio
is relative to \texttt{pure\_as} in the same asset-window pair.}
\label{tab:stress-backtest}
\end{table}

The stress-window result confirms that the HJB effect is not uniformly
positive. It improves mean final equity in eight of twelve asset-window pairs
and reduces inventory RMS in ETH and BTC, but it loses in ETH high-funding and
high-volatility windows and in BTC high-volatility and calm windows. SOL remains
strong in mean equity but again relies on much larger inventory exposure.

\section{Discussion}

The present result supports a specific and limited claim. A
validation-selected finite-difference HJB that treats funding as a state
variable improves mean ETH/BTC holdout performance relative to classical
funding-unaware AS while lowering inventory RMS. This is true under the
baseline \texttt{volume\_minute} effective-fill calibration and under the
alternative \texttt{minute\_hit} calibration. The result is therefore not an
artifact of one particular official-fill counting rule. It is also not a claim
of uniform pathwise dominance: BTC win rates are modest, and the confidence
intervals for paired deltas are wide enough that the claim should remain about
mean performance and risk exposure, not deterministic superiority.

The SOL result is economically large but less clean. The HJB earns more than
unscaled AS in every seed, but it does so with much larger inventory RMS. A
risk-scaled AS diagnostic earns substantially more than the HJB on SOL. That
comparison changes the interpretation. On SOL, the HJB is not a Pareto
improvement over AS; it is a different risk-return point that takes more
inventory exposure in a market where such exposure is rewarded by the proxy
simulator. Reporting this honestly is important because otherwise the backtest
would confuse compensation for inventory risk with a funding-aware control
advantage.

The main theoretical limitation is the funding law. Pure OU is retained because
it gives a tractable diffusion state, a closed-form transition density for
calibration, and a monotone finite-difference HJB. The data, however, favor
OU-plus-jump transitions. A natural next model is a jump HJB in which funding
shocks enter through a nonlocal generator. Another practical alternative is a
regime-switching funding model, where the HJB is solved across a small number
of funding regimes with different local mean-reversion and jump-risk
parameters. Either extension would make the model closer to the observed
funding residuals.\footnote{As a preliminary cross-venue check, a six-month
Binance ETHUSDT funding calibration over 4 November 2025 through 3 May 2026
gives an OU half-life of 7.96 hours and an OU-plus-jump train likelihood gain
of 32.42. This is directionally consistent with the Hyperliquid ETH evidence
that funding is mean reverting and jump diagnostics improve likelihood,
although it is not a full Binance backtest.}

The main empirical limitation is execution. The fill model is calibrated from
official Hyperliquid crossed fills and L2 quotes, and the ETH/BTC result
survives two official-fill proxy calibrations. Nevertheless, neither proxy
knows the market maker's queue position, latency, maker priority, cancellation
behavior, or adverse selection conditional on being filled. The simulator is
best read as a controlled policy comparison under common execution assumptions.
Before a stronger trading claim, the next robustness layer should add a simple
adverse-selection or latency cost, or build a queue-position simulator if the
raw data support it.

Relative to perpetual liquidation work, the contribution is the endogenous
creation of inventory through bid and ask quoting. A liquidation agent reduces
or reshapes an existing position; the market maker in this paper chooses quote
offsets that determine when inventory is acquired and when it is unwound.
Relative to AMM predictable-loss work, the mechanism is also different. The
policy is not a constant-product or concentrated-liquidity range. It is an
order-book-style feedback rule derived from an inventory--funding HJB.

The most useful next paper version would make two changes. First, it would move
the OU-plus-jump funding evidence into the control problem, either through a
nonlocal HJB or through a regime-switching approximation. Second, it would add
execution frictions directly to the simulator. If the ETH/BTC result survives
both changes, the paper would have a substantially stronger claim: funding is
not only a statistically meaningful state variable, but a practically useful
state variable for inventory-aware liquidity provision.

\appendix
\section{Linear-Quadratic Prototype and Numerical Policy}

Before switching to the finite-difference HJB in Section~\ref{sec:hjb}, the
repo used a projected linear-quadratic approximation with coefficients
$(a_0,\ldots,a_5)$ and an inventory-funding cross term. The prototype assumed
\[
\theta(t,q,f)=
a_0(t)+a_1(t)f+a_2(t)f^2+a_3(t)q^2+a_4(t)qf+a_5(t)q.
\]
The term $a_4(t)qf$ is the main structural intuition: the marginal value of
inventory should depend on the current funding state. A positive funding state
makes long inventory less attractive and short inventory more attractive; a
negative funding state reverses that ranking.

This approximation is useful for intuition but is not the final strategy. Once
the exponential arrival Hamiltonian is kept in the control equation, projecting
the full Hamiltonian back onto a quadratic basis requires additional
approximations. Those approximations were not robust enough in the holdout
experiments. The final policy therefore solves the discrete HJB directly and
recovers quotes from neighboring inventory values.

The numerical policy used in the paper can be summarized as follows. First,
choose grids $(t_i,q_j,f_\ell)$ and initialize
$\theta_{N,j,\ell}=-\alpha q_j^2$. Second, solve backward with the monotone
birth--death funding operator and arrival Hamiltonians. Third, interpolate
$\theta(t,q,f)$ in time and funding at each simulated quote time. Finally,
compute bid and ask offsets from $1/k-\Delta_q\theta$ and apply quote floors
and inventory-boundary rules.
The implementation uses Numba-accelerated bilinear interpolation for quote
recovery. This is computational rather than theoretical, but it matters for the
empirical design: the final 100-seed cross-asset robustness runs are feasible
because the HJB lookup is constant-time rather than a scan over the full time
grid.

The appendix is therefore not a closed-form solution. It records the
linear-quadratic intuition and clarifies why the reported empirical strategy is
the finite-difference HJB rather than the Riccati prototype.

\bibliographystyle{plain}
\bibliography{references}

\end{document}